\documentclass[namedreferences, final, b5paper, noid]{solarphysics}

\usepackage[optionalrh]{spr-sola-addons} 
\usepackage{graphicx}        
\usepackage{color}           
\usepackage{url}             
\usepackage[pdfborder={0 0 0 },urlcolor=blue]{hyperref}  

\begin{document}

\ifx \doiurl  \undefined \def \doiurl#1{\href{http://dx.doi.org/#1}{\url{#1}}}\fi
\ifx \adsurl  \undefined \def \adsurl#1{\href{http://adsabs.harvard.edu/abs/#1}{\url{#1}}}\fi

\renewcommand\UrlFont{\sf}            
\newcommand{\BibTeX}{\textsc{Bib}\TeX}
\newcommand{\etal}{{\it et al.}}

\newcommand{\deriv}[2]{\frac{{\mathrm d} #1}{{\mathrm d} #2}}
\newcommand{\rmd}{ {\ \mathrm d} }
\renewcommand{\vec}[1]{ {\mathbf #1} }
\newcommand{\uvec}[1]{ \hat{\mathbf #1} }
\newcommand{\pder}[2]{ \f{\partial #1}{\partial #2} }
\newcommand{\grad}{ {\bf \nabla } }
\newcommand{\curl}{ {\bf \nabla} \times}
\newcommand{\vol}{ {\mathcal V} }
\newcommand{\bndry}{ {\mathcal S} }
\newcommand{\dv}{~{\mathrm d}^3 x}
\newcommand{\da}{~{\mathrm d}^2 x}
\newcommand{\dl}{~{\mathrm d} l}
\newcommand{\dt}{~{\mathrm d}t}
\newcommand{\intv}{\int_{\vol}^{}}
\newcommand{\inta}{\int_{\bndry}^{}}
\newcommand{\avec}{ \vec A}
\newcommand{\ap}{ \vec A_p}
\newcommand{\bb}{ \vec B}
\newcommand{\jj}{ \vec j}
\newcommand{\rr}{ \vec r}
\newcommand{\xx}{ \vec x}

\newcommand{\adv}{    {\it Adv. Space Res.}}
\newcommand{\annG}{   {\it Annales Geophysicae}}
\newcommand{\aap}{    {\it Astron. Astrophys.}}
\newcommand{\aaps}{   {\it Astron. Astrophys. Suppl.}}
\newcommand{\aapr}{   {\it Astron. Astrophys. Rev.}}
\newcommand{\ag}{     {\it Ann. Geophys.}}
\newcommand{\aj}{     {\it Astron. J.}}
\newcommand{\apj}{    {\it Astrophys. J.}}
\newcommand{\apjl}{    {\it Astrophys. J. Lett.}}
\newcommand{\apss}{   {\it Astrophys. Space Sci.}}
\newcommand{\cjaa}{   {\it Chin. J. Astron. Astrophys.}}
\newcommand{\gafd}{   {\it Geophys. Astrophys. Fluid Dyn.}}
\newcommand{\grl}{    {\it Geophys. Res. Lett.}}
\newcommand{\ijga}{   {\it Int. J. Geomag. Aeron.}}
\newcommand{\jastp}{  {\it J. Atmos. Solar Terr. Phys.}}
\newcommand{\jgr}{    {\it J. Geophys. Res.}}
\newcommand{\mnras}{  {\it Mon. Not. Roy. Astron. Soc.}}
\newcommand{\nat}{    {\it Nature}}
\newcommand{\pasp}{   {\it Pub. Astron. Soc. Pac.}}
\newcommand{\pasj}{   {\it Pub. Astron. Soc. Japan}}
\newcommand{\pre}{    {\it Phys. Rev. E}}
\newcommand{\solphys}{{\it Solar Phys.}}
\newcommand{\sovast}{ {\it Sov. Astron.}}
\newcommand{\ssr}{    {\it Space Sci. Rev.}}

\begin{article}

\begin{opening}

\title{The SWAP EUV Imaging Telescope Part I: Instrument Overview and Pre-Flight Testing}

\author{D.B.~\surname{Seaton}$^{1}$\sep
        D.~\surname{Berghmans}$^{1}$\sep
        B.~\surname{Nicula}$^{1}$\sep
        J.-P.~\surname{Halain}$^{2}$\sep
        A.~\surname{De Groof}$^{1,3}$\sep
        T.~\surname{Thibert}$^{2}$\sep
        D.S.~\surname{Bloomfield}$^{4}$\sep
        C.L.~\surname{Raftery}$^{4, 5}$\sep
        P.T.~\surname{Gallagher}$^{4}$\sep
        F.~\surname{Auch\`ere}$^{6}$\sep
        J.-M.~\surname{Defise}$^{7}$\sep
        E.~\surname{D'Huys}$^{1}$\sep
        J.-H.~\surname{Lecat}$^{2}$\sep
        E.~\surname{Mazy}$^{2}$\sep
        P.~\surname{Rochus}$^{2}$\sep
        L.~\surname{Rossi}$^{2}$\sep
        U.~\surname{Sch\"uhle}$^{8}$\sep
        V.~\surname{Slemzin}$^{9}$\sep
        M.S.~\surname{Yalim}$^{10}$\sep
        J.~\surname{Zender}$^{3}$
       }
\runningauthor{D.B.~Seaton {\it et al.}}
\runningtitle{The SWAP EUV Imaging Telescope\,---\,I}

   \institute{$^{1}$ Royal Observatory of Belgium, Avenue Circulaire 3, 1180 Brussels, Belgium
                     email: \url{dseaton@oma.be}\\ 
              $^{2}$ Centre Spatial de Li\`ege, Universit\'e de Li\`ege, Li\`ege Science Park, 4013 Angleur, Belgium \\
              $^{3}$ ESA Directorate of Scientific and Robotic Exploration, ESTEC, Noordwijk, The Netherlands \\
              $^{4}$ Astrophysics Research Group, School of Physics, Trinity College Dublin, Dublin 2, Ireland \\
              $^{5}$ Space Sciences Lab, UC Berkeley, 7 Gauss Way, Berkeley, CA 94720-7450, USA \\
              $^{6}$ Institut d'Astrophysique Spatiale, CNRS, Univ. Paris-Sud 11, France  \\
              $^{7}$ Institut d'Astrophysique et de G\'eophysique, Universit\'e de Li\`ege, 4000 Li\`ege, Belgium \\
              $^{8}$ Max-Planck Institut f\"ur Sonnensystemforschung, Katlenburg-Lindau, Germany \\
              $^{9}$ P.N. Lebedev Physical Institute, Leninsky Pr., 53, Moscow, 119991, Russia \\
              $^{10}$ KULeuven, Center for Plasma Astrophysics, Celestijnenlaan 200b - bus 2400, 3001 Heverlee, Belgium \\
             }

\begin{abstract}
The \emph{Sun Watcher with Active Pixels and Image Processing} (SWAP) is an EUV solar telescope on board ESA's \emph{Project for Onboard Autonomy~2} (PROBA2) mission launched on 2~November~2009. SWAP has a spectral  bandpass centered on 17.4~nm and provides images of the low solar corona over  a 54$\times$54~arcmin field-of-view with 3.2~arcsec pixels and an imaging cadence of about two~minutes.  SWAP is designed to monitor all space-weather-relevant events and features in the low solar corona. Given the limited resources of the PROBA2 microsatellite,  the SWAP telescope is designed with various innovative technologies, including an off-axis optical design and a CMOS--APS detector. This article provides reference documentation for users of the SWAP image data. 
\end{abstract}
\keywords{Instrumentation and Data Management; Corona, Structures}
\end{opening}

\setcounter{page}{1}
\pagenumbering{arabic}

\section{Introduction}
     \label{Introduction} 

The \emph{Project for On Board Autonomy} (PROBA) missions are a series of microsatellites launched by the European Space Agency (ESA) and intended to provide an in-orbit test platform for new technologies. PROBA1 was launched in 2001 and carried instruments primarily intended for Earth observation.  PROBA2 \cite{TI_PROBA2_Platform_Santandrea} was inserted into a Sun-synchronous polar orbit at an altitude of approximately 720~km. This orbit guarantees nearly uninterrupted solar viewing through most of the year. The nominal two-year mission started with the 2~November~2009 launch but was subsequently extended until the end 2012.

The primary mission goal of PROBA2 is to perform an in-flight demonstration of a series of new spacecraft technologies. The secondary mission goal is the exploitation of the payload of scientific instruments consisting of two Sun-sensing instruments, the \emph{Sun Watcher with Active Pixel Sensor and Image Processing} (SWAP), the main subject of this article, and the \emph{Lyman-Alpha Radiometer} (LYRA: \opencite{Hochedez06}, \opencite{TI_PROBA2_LYRAinstrumentPaper_Dominique}).  Also on board are two instruments intended to measure {\it in-situ} properties of the magnetosphere: the \emph{Thermal Plasma Measurement Unit} (TPMU) and the \emph{Dual-Segmented Langmuir Probe} (DSLP).

\begin{sloppypar}
As a microsatellite, the PROBA2 platform imposes severe limitations on the payload resources such as volume, mass, and power. Thus SWAP was designed as a miniaturized coronal imager, capable of meeting these limitations, but also capable of extending the successful ``CME watch'' program that began with data from the \emph{Extreme-Ultraviolet Imaging Telescope} (EIT) on board the joint ESA--NASA \emph{Solar and Heliospheric Observatory} (SOHO) mission \cite{Delaboud95}. By improving the imaging cadence by an order of magnitude over that of EIT, SWAP can track the evolution of all events in the low solar corona that are relevant for space weather, including flares, CMEs, EUV waves, and EUV dimmings. In addition, SWAP continuously images solar features such as coronal holes and active regions, the locations of which are essential data for space weather forecasters. As a result, while SWAP provides useful data for both scientific analysis and space weather forecasting in the present, the lessons learned during the development of SWAP are being incorporated into a future generation of space weather monitors as well. One example is the \href{http://sidc.be/esio/}{\emph{EUVI Solar Imager for Operations} (ESIO)}, a potential building block of the ESA Space Situational Awareness Program.
\end{sloppypar}

Compared to more resource-rich contemporary EUV imagers such as the \emph{Extreme Ultraviolet Imagers} (EUVI), part of the \emph{Sun Earth Connection Coronal and Heliospheric Investigation} (SECCHI) packages on the twin \emph{Solar Terrestrial Relations Observatory} (STEREO) spacecraft \cite{2008SSRv..136...67H}, and the \emph{Atmospheric Imaging Assembly} (AIA) instrument on board the \emph{Solar Dynamics Observatory} (SDO; \opencite{2012SoPh..275...17L}), SWAP offers only modest temporal and spatial resolution. However, SWAP's design and operational strategy result in some unique capabilities compared to other EUV solar imagers.

In particular, SWAP images are useful because they provide the largest EUV field-of-view available from Earth orbit (54~arcmin) and SWAP's complementary metal-oxide-semiconductor active-pixel system (CMOS--APS) detector does not bloom significantly during bright flares, when nearly all other detectors do so. In addition, the agility feature of PROBA2 permits SWAP to off-point by up to one~degree, allowing imaging of extended coronal features and tracking of CMEs as they propagate away from the Sun.

In this article we provide a reference description of the SWAP instrument, the results of several pre-flight instrumental tests, and an overview of spacecraft operations that are likely to be particularly relevant to data users in the solar physics and space weather communities. The second part of this article, \inlinecite{TI_PROBA2_SWAPcalibrationPaper_Halain}, discusses in-flight calibration and performance of the SWAP instrument. \inlinecite{Berghmans06} discuss the early development of the SWAP instrument, while many of the technological improvements associated with the development of SWAP are reviewed by \inlinecite{Defise07}.

The material presented here is a synthesis of technical specifications that appear in the data sheets for the several instrument and spacecraft subsystems, the instrument design itself, and from data collected during two calibration campaigns at the \href{http://www.ptb.de/mls/aufgaben/bessylab.html}{PTB/BESSY} facilities in Berlin, Germany, in February 2007 and July 2008. These calibration campaigns included tests of individual parts of the SWAP instrument as well as the complete SWAP optical path and detector. (During the July 2008 campaign we conducted these tests using the flight models of the instrument proximity electronics and instrument interface unit.)

In Section~\ref{Overview} we give an overview of the instrument and introduce the main subsystems of the telescope. In the following sections we focus on the main functional aspects of the SWAP telescope that are relevant for SWAP data users: spectral selection (Section~\ref{Spectral}), imaging (Section~\ref{Imaging}), signal recording (Section~\ref{SignalRecording}), and on board data processing (Section~\ref{OnboardDataProcessing}).  After that we discuss the limitations and opportunities provided by SWAP's operational strategy (Section~\ref{Operations}) as well as the SWAP data products and analysis software that are available to users (Section~\ref{DataProducts}) through the PROBA2 Science Center. Finally, we make some concluding remarks in Section~\ref{Conclusions}. For quick access to reference material, a SWAP data sheet is provided in an appendix.

Additional articles in this same \emph{Solar Physics Topical Issue} concentrate on the PROBA2 platform \cite{TI_PROBA2_Platform_Santandrea}, on the LYRA ``sister'' instrument \cite{TI_PROBA2_LYRAinstrumentPaper_Dominique}, and on the management of SWAP and LYRA operations from the PROBA2 Science Center \cite{TI_PROBA2_P2SC_Zender}. 

\section{Instrument Overview}
     \label{Overview} 

\begin{figure}[ht]
\centering
\includegraphics[scale=0.75, clip = true]{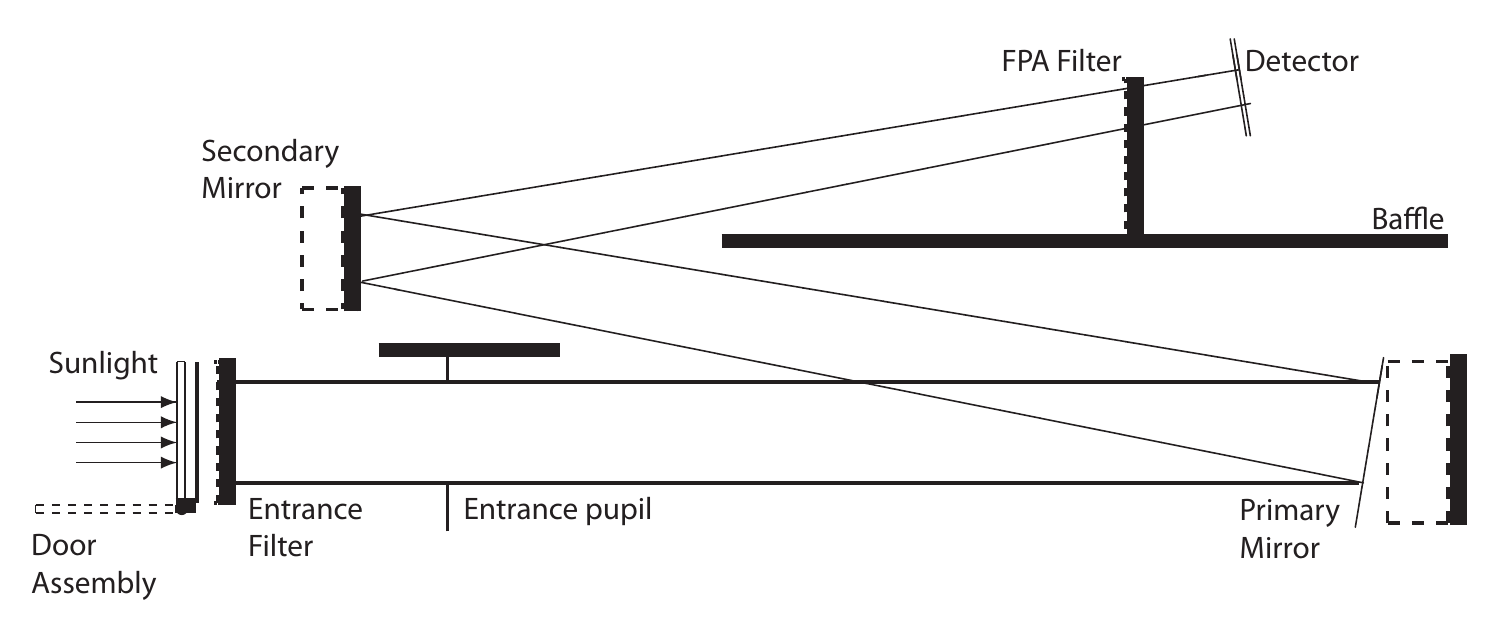} 
\caption{Schematic overview of the SWAP optical path, which is folded into SWAP's $565\times150\times125$~mm case. }\label{fig:swap-schematic}
\end{figure}

The design of the SWAP telescope body was driven by the limited spacecraft dimensions, the available mass budget, and the relatively harsh conditions to which the spacecraft was subjected during launch. A novel two-mirror off-axis Ritchey--Chr\'etien scheme was chosen (Figure~\ref{fig:swap-schematic}) for SWAP because it minimizes the telescope length and allows for a simple and efficient internal baffling system. The absence of a central obstruction (such as, for example, EIT's secondary mirror) allows for a smaller aperture\,---\,SWAP's aperture is roughly 33~mm in diameter\,---\,and hence smaller and lighter mirrors. An additional advantage is that the smaller aperture could use smaller-diameter aluminum-foil filters\,---\,used to reject visible and infrared light from the telescope\,---\,which reduced the risk of damage during launch on board the Russian \emph{Rockot} launch system that injected PROBA2 into orbit.

SWAP's two mirrors are made of Zerodur\texttrademark, polished to a micro-roughness below 0.5~nm to limit the scattering of EUV light that can degrade the off-disk part of the image. Two aluminum filters, one at the entrance to the telescope and one located in front of the focal plane assembly (FPA), limit visible and infrared light contamination in the telescope.

The optical elements of SWAP are mounted on an Invar\texttrademark optical bench that provides mechanical stability and thermal invariance (less than 50-$\mu\textrm{m}$ between the two mirrors: \opencite{Defise07}). This thermal invariance is needed to keep the telescope in focus despite the wide thermal range of the environment in which SWAP is expected to operate. The SWAP telescope is designed to operate in a temperature range between $-32~^{\circ}\textrm{C}$ and $+20~^{\circ}\textrm{C}$, although nominal operating temperatures are closer to $0^{\circ}\textrm{C}$. (In fact, SWAP can survive in temperatures between $-40~^{\circ}\textrm{C}$ and $+60~^{\circ}\textrm{C}$, but cannot actually operate across that entire range.) Thermal control over SWAP itself is strictly passive: the bench is thermally driven by the platform itself. The detector itself is connected via a thermal link (a ``cold finger'') to an external radiator. All other thermal interaction is damped by a multilayer insulation blanket wrapped around the instrument. 

The instrument is encased in a lightweight Invar housing and protected from contamination during on-ground activities and launch phase by a one-shot door that was opened once PROBA2 reached orbit. The front door mechanism, which is directly exposed to the Sun, is painted white to limit heat absorption.

Besides the proximity electronics that are used to read-out and control the detector and are integrated in the telescope body, SWAP uses two external electronic boxes. The Instrument Interface Unit (IIU) serves as the power conditioning unit, and is shared with LYRA. The Memory Compression and Packetization Module (MCPM) board is the data-processing unit, which is integrated in the data-processing subsystems of the PROBA2 platform electronics (the Advanced Data and Power Management System, ADPMS).

The resulting dimensions of SWAP are $565\times150\times125$~mm. Its mass is approximately 11~kg for the telescope, the IIU, and two MCPM boards. SWAP's peak power consumption is just 2.6~W, including all power necessary for both the detector operations and proximity electronics.

\section{Spectral Performance}
    \label{Spectral} 

SWAP was designed to image the Sun in a bandpass centered on 17.4~nm. This bandpass contains the brightest coronal emission lines in the EUV spectrum, and simultaneously provides satisfactory response to emission from the features and events that serve as the drivers of space weather. Thus the selected bandpass represents an excellent compromise between overall instrument sensitivity and sensitivity to the features associated with SWAP's science objectives.

Spectral selection is achieved by a combination of aluminium-foil filters and multilayer coatings on the mirrors.  The front aluminium filter, which has a diamater of 38~mm, reflects both solar heat load and visible light at the entrance of the telescope. Although the presence of a focal-plane filter means that a pinhole in the front filter would not be catastrophic, any major failure of the front filter due to venting, acceleration, or vibrations during launch would have rendered SWAP images essentially unusable. Thus the filter's design represents a compromise between optimal optical performance and the mechanical strength necessary to survive the spacecraft's ascent. In order to strike the necessary balance, the filter is composed of a 150-nm aluminium foil that was deposited on a 50-nm polyimide film and reinforced with a nickel mesh grid with a spacing of 1.270~mm and grid width 0.045~mm.

The use of a nickel grid in the front filter is known to produce a diffraction pattern in the images (see, for example, \opencite{DeForestTrace2009}; \opencite{2011ApJ...743L..27R}). However, given the characteristics of the nickel grid selected for the filter (which yields 90\,\% transmission), the secondary peaks are weak in intensity and only diffracted over about the distance of one pixel. Thus the effect of this grid is only a moderate broadening of the effective point-spread-function (PSF) beyond the PSF due to optical elements in the telescope other than the filter grid.

The ratio of solar visible light to EUV is so large that a second filter is also required to remove unwanted visible light. This filter, which has a diameter of 28~mm, is located near the FPA at the end of the telescope's optical path to eliminate any residual visible light and to protect against any possible light leaks in the front housing or the entrance filter itself. The FPA filter uses a more traditional design than the front filter: a 150-nm thick aluminium foil supported by a nickel grid with a grid spacing of 0.363~mm and grid widths of 0.037~mm (yielding 80\,\% transmission). The nickel grid produces small shadow artifacts, but these were modeled to be less than 1\,\% of total signal and are essentially invisible in normal SWAP images. (These artifacts are occasionally visible in running-difference movies, but their overall effect on image quality is negligible.) 

EUV reflectivity on the mirrors and spectral filtering around the Fe~\textsc{ix}/\textsc{x} emission lines (17.4~nm) is the result of EUV multilayer coatings on the two mirrors. Both SWAP's primary (M1) and secondary (M2) mirrors were coated with Mo/Si multilayers using an ion beam sputtering process that was first developed for the four channels of the EUVI telescopes on the STEREO/SECCHI instrument \cite{Ravet04}.

We measured the EUV reflectivity of SWAP's mirrors, for both calibration and witness samples, at the operational wavelength at two separate facilities. One set of tests were performed at the synchrotron calibration facility of Physikalisch-Technische Bundesanstalt (PTB) at the Berlin Electron Storage Ring for Synchrotron Radiation (BESSY II) in Berlin, while other tests used the EUV plasma source reflectometer Centrale de d'Elaboration et de Metrologie des Optiques X-Orsay (CEMOX). We evaluated the multilayer performance using a cross-analysis of the data obtained at both facilities as well as additional measurements from the BEAR line of the \href{http://www.elettra.trieste.it}{ELETTRA} synchrotron facility in Trieste, Italy. The results of this performance evaluation appear in Table~\ref{TableMultilayers}.

\begin{table}
\caption{Reflection characteristics of SWAP flight model mirrors FM1 and FM2. The bandwidth specified is the full width at half-max measured using the synchrotron beam at BESSY.}
\label{TableMultilayers}
	\begin{tabular}{ c   c   c   c }
          \hline
           & $\lambda$ max [nm] 	& 	Reflectivity [\%] 	& Bandwidth [nm]  \\
          \hline
          \textbf{FM1} & $17.47 \pm 0.1 $ & $40.80 \pm 2.00$ & 1.3 \\
          \textbf{FM2} & $17.62 \pm 0.1 $ & $39.50 \pm 2.00$ & 1.4 \\
          \hline
	\end{tabular}
\end{table}

\subsection{End-to-End Throughput}


During the second of our two BESSY calibration campaigns, we characterized the end-to-end throughput and response function of the SWAP imager, including the effect of all filters, mirrors, and detector, by measuring the variation in detected beam signal strength over a range of wavelengths near SWAP's expected passband. To do this, we varied the wavelength of the input beam from 16.50~nm to 20.00~nm using increments of 0.1~nm. For each increment we acquired images of the beam spot that could be compared to the strength of the input beam in order to find the throughput at each particular wavelength.  However, because the synchrotron source at BESSY degrades over time, the beam strength slowly decreased as the throughput test unfolded. So it was also necessary to monitor the strength of the input beam before it entered the instrument throughout the test. We did this by inserting a small photodiode into the beam at regular intervals throughout the test and recording the changing beam strength.

After the resulting images are dark-subtracted and normalized for integration time, we can measure the total detected signal strength by taking the sum of all counts across the entire detector.
\begin{equation}
\label{eqn:counts_cal}
I_{\mathrm{det}} =  \frac{1}{t} \displaystyle\sum_{\textrm{\scriptsize Pixels}}C \hspace{2.2cm} [\textrm{DN}\ \textrm{s}^{-1}], \end{equation}
where $t$ is the integration time, and $C$ represents the signal (voltage in DN) collected by each detector pixel. Since this value is proportional to the photon flux that actually reaches the detector, the instrumental throughput is simply the ratio of the detected signal to the incident signal, $I_{\mathrm{beam}}$ (which has units of $\textrm{photons } \textrm{s}^{-1}$), so

\begin{equation}
\label{eqn:cal}
R_{\mathrm{throughput}} =  \frac{I_{\mathrm{det}}}{I_{\mathrm{beam}}}  \hspace{2.2cm}   [\textrm{DN phot}^{-1}].
\end{equation} 

The fraction of total solar illumination observed by any single pixel depends on both the aperture size and solid angle covered by each pixel, and is given by the product of the area of the apterture ($\textrm{cm}^{2}$) and the steradian coverage of one pixel ($\textrm{str}\ \textrm{pix}^{-1}$). Thus the instrumental response is given by the ratio, $R$, of total measured signal to total input flux times this geometric factor. The procedure used to determine the instrument throughput is discussed in more detail in Chapter~6 of \inlinecite{RafteryThesis}.

\begin{figure}[ht]
\centering
\includegraphics[scale=0.75, clip = true]{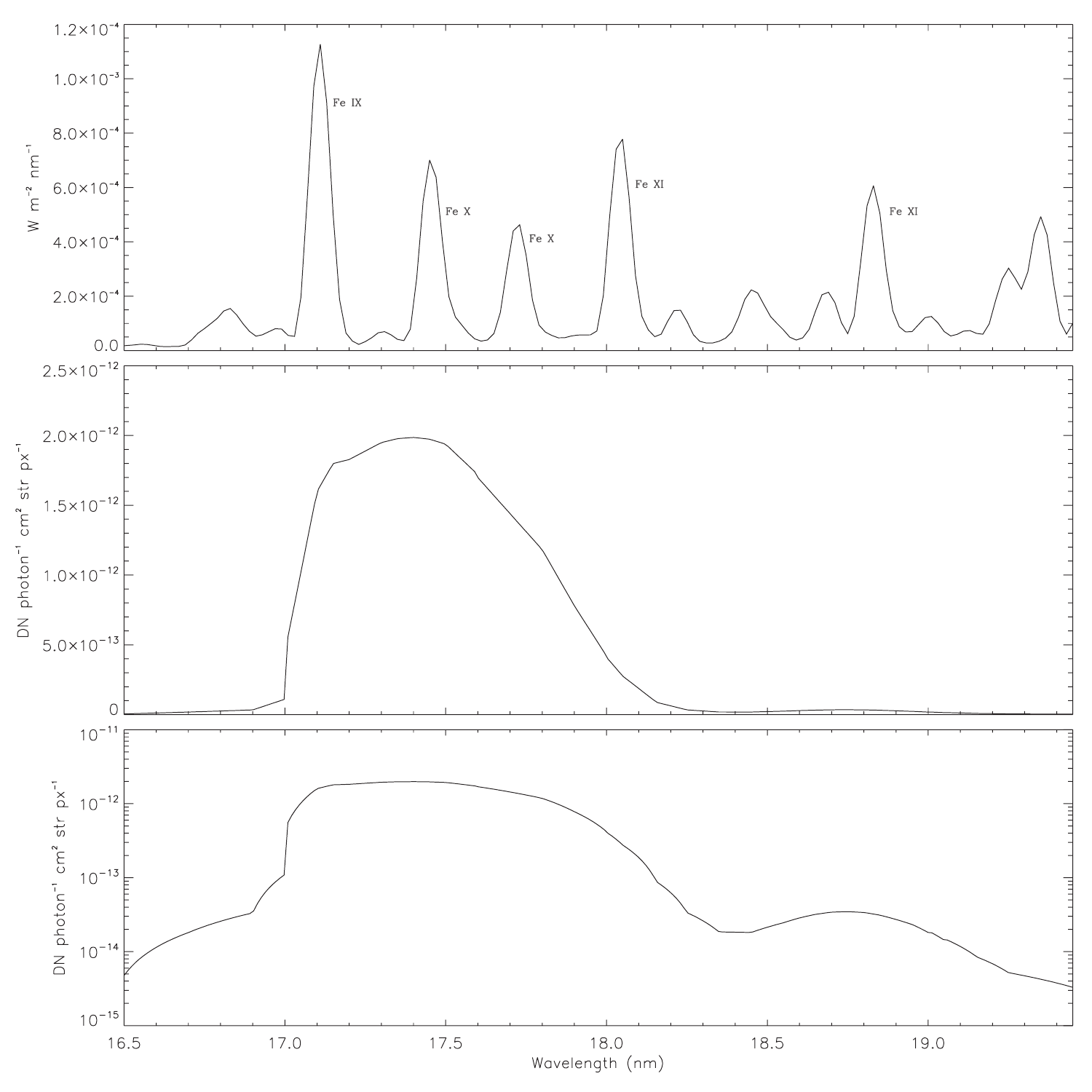} 
\caption{SWAP response as a function of wavelength. The top panel shows a representative solar spectrum indicating primary emission lines in the region of SWAP's sensitivity, measured by the Extreme Ultraviolet Variability Experiment (EVE) on SDO on 1~January~2012. The response function itself is plotted on linear (middle panel) and log (bottom panel) scales to show both the region of peak transmission near 17.4~nm and the secondary transmission peak at longer wavelengths.}
\label{fig:swap_resp}
\end{figure}

It is worth noting that the preliminary response function shown in \inlinecite{RafteryThesis} was not corrected for the shadowing effect of the filter grid on the highly collimated beam used for the test at BESSY. A careful analysis of the response, including cross-calibration with spectral observations from the EUV Variability Experiment (EVE) on SDO discussed in \inlinecite{TI_PROBA2_SWAPcalibrationPaper_Halain}, revealed that because of shadowing due to the beam position with respect to the filter grids\,---\,which can modulate the beam strength by nearly 20\,\%\,---\,the response function required a correction of about 12\,\%. (A complete discussion of the grid shadowing effect is included in Section~\ref{sec:filter_scan}.)  We have, therefore, adjusted the instrument response function shown in Figure~\ref{fig:swap_resp} from its first appearance in \inlinecite{RafteryThesis} to values more consistent with those measured during SWAP's in-flight calibration campaigns. (Note that the measured response function itself and the telescope properties reported in the appendix are also included in the SWAP software packages distributed though \textsf{SolarSoft}.)

\inlinecite{Raftery2012_Imagers} include a detailed comparison of SWAP's instrumental response to the response functions of the corresponding EUV channels of similar imagers, including SDO/AIA, STEREO/EUVI, SOHO/EIT, and TRACE.

\subsection{Radiometric Calibration}
\label{sec:rad_cal}

SWAP's CMOS detector's response to light is known to be slightly nonlinear, and early tests on the detector showed that the linearity of its response depends on both the inherent brightness of the image source and the integration time of the each acquired image itself. In order to set some overall limits on the linearity of the instrumental response and disentangle the two contributions to nonlinearity as much as possible, we performed a series of measurements intended to characterize the SWAP's overall radiometric performance. We performed two separate tests using the BESSY beam.  In the first set of tests, we obtained images of a beam of constant intensity with a variety of integration times, while in the second we used a fixed integration time and varied the beam strength itself, which we did by adjusting the size of the aperture that regulates the overall intensity of the input beam. During both of these test sequences the BESSY beam was fixed at 17.4~nm, corresponding to the wavelength of SWAP's maximum sensitivity. 

During the variable integration time sequence acquired images with integration times between 1~second and 22~seconds and the beam intensity tuned such that there was minimal saturation even in the images with the longest integration time. Because each pixel in the SWAP CMOS detector has its own readout electronics\,---\,and, therefore, its own radiometric response\,---\,we measured the response in several locations on the detector to confirm that the radiometric performance of the detector as a whole was not spatially variable. These measurements were also performed using both of the detector's two readout modes. (The different readout modes of SWAP's CMOS detector are described in detail in Section~\ref{SignalRecording}.)

One important conclusion of these tests was that the radiometric performance of the instrument depends strongly on the readout mode of the detector. In particular, the use of Double Sampling (DS) mode can lead to the appearance of ``image lag'' artifacts in the images. These artifacts can appear in image sequences where pixels that are illuminated in one image are not completely reset before the beginning of a subsequent image. This effect is discussed in detail in \inlinecite{Degroof08}. Since this effect is almost completely mitigated by the use of SWAP's other readout mode, Correlated Double Sampling (CDS) mode, and nearly all SWAP images are obtained in this mode, we will not discuss it in detail here.

Another important conclusion of both radiometric tests is that SWAP responds linearly to within approximately 5\,\% to all input, independent of integration time or incoming intensity. The detector is most linear in the middle of its sensitivity range and least linear near its saturation level. Although there is some nonlinearity associated with very low signal levels, carefully configuring the detector and proximity electronics in SWAP largely limits the nonlinearity to nearly saturated pixels. (The electronic configuration of the detector is discussed in more detail in Section~\ref{SignalRecording}.)

During the second BESSY campaign we also measured the variation in detector response as a function of temperature. Although increased temperature indeed leads to increased noise in SWAP images, we found that temperature has a negligible effect on the instrument response itself. Since much of the additional noise due to increased temperature is dark current, which can be removed during the image calibration process, temperature also has a negligible effect on detector response.

It is also worth pointing out that this campaign revealed several thousand pixels in SWAP's CMOS detector that are either nonfunctional or have a response well outside the nominal performance of the instrument. These anomalous pixels are excluded from SWAP images using an on board map of bad pixels (see Section~\ref{OnboardDataProcessing}). Most pixels nevertheless perform within a few percent of the detector mean, are insensitive to temperature, and are linear in all but a small portion of their dynamic range. Thus we conclude that the instrumental radiometric performance is good to within approximately 5\,\%.

\section{Telescope Performance}
     \label{Imaging} 

Besides SWAP's sensitivity to various spectral lines in its passband, SWAP images depend on two major effects: first, the optical properties of the SWAP telescope itself and, second, the performance and configuration of SWAP's CMOS--APS detector. Here we discuss the physical properties of the instrument itself; we discuss SWAP's electronics and detector in the next section. In particular, we will focus here on the effect of the instrument design on the appearance of the images, most notably the presence of stray light due to reflections from structures inside the telescope itself and the influence of the filters themselves on image quality. 

\subsection{Stray Light}

A major concern for SWAP is the presence of stray light, especially during observations of the extended off-limb parts of the corona that are outside the fields-of-view of most other imagers, but where the intrinsic EUV brightness of the corona is also relatively poor.  During nominal operations the main potential source of stray light in SWAP images is visible-light leaks, since the number of visible photons from the Sun is many orders of magnitude greater than the number of EUV photons in the SWAP bandpass.  There is no evidence for such leaks in any SWAP images obtained since launch. 

\begin{figure}[ht]
\centering
\includegraphics[scale=0.6, clip = true]{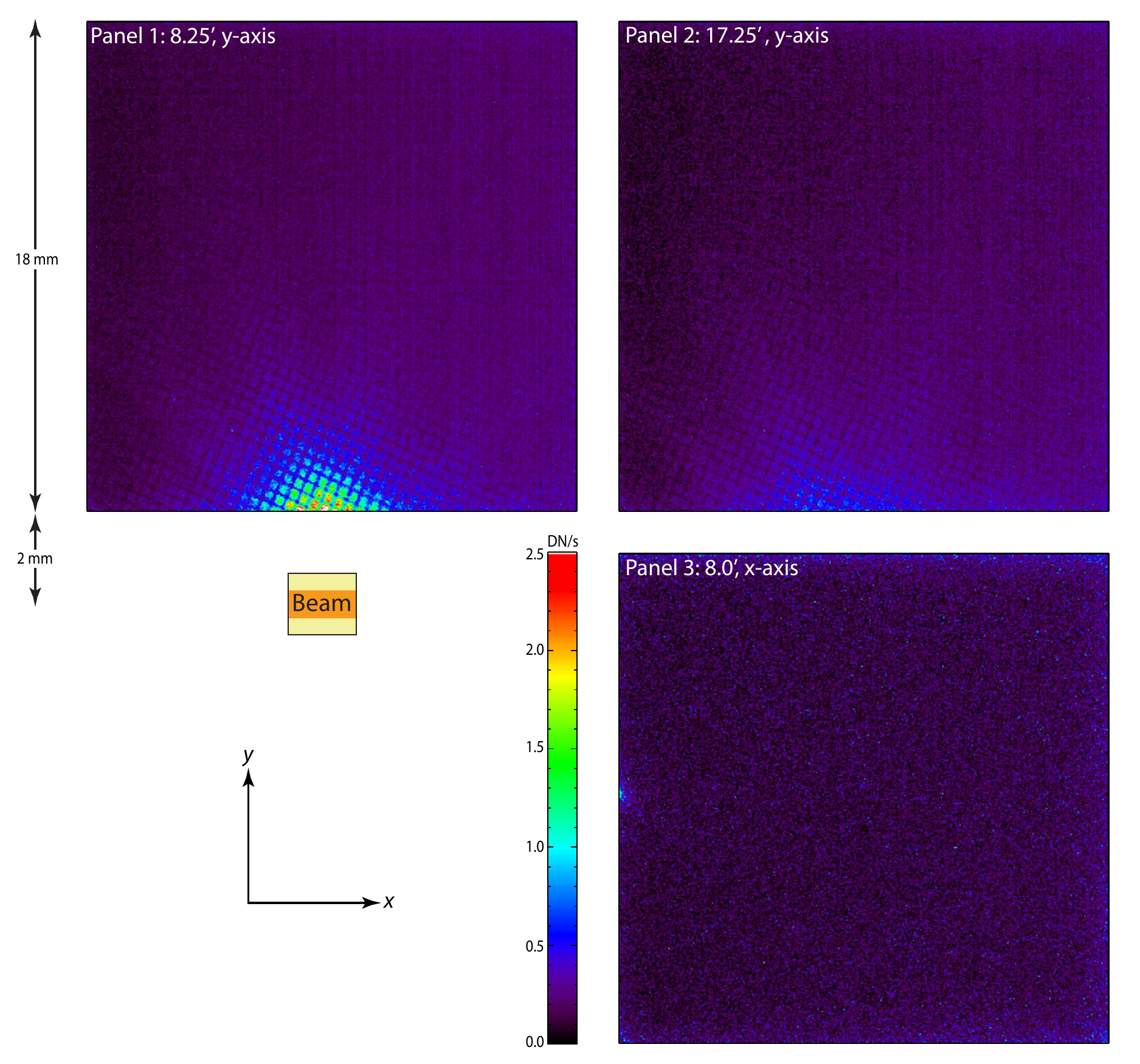} 
\caption{Stray light for three different beam positions. Panels~1 and 2 show stray light for beam displacements of 8.25 and 17.25~arcmin in the $y$-direction; Panel~3 shows stray light for a beam displacement of 8~arcmin in the $x$-direction. The schematic shows how far the beam was displaced in the $y$-direction to generate the image shown in Panel~1. The coordinate axis is indicated in the lower left.}
\label{fig:straylight_images}
\end{figure}

However, EUV contributions from the on-disk corona can also be scattered inside the telescope even when the Sun is fully outside of the detector's field-of-view, leading to the possibility of stray light appearing in both Sun-centered and off-pointed images. In order to characterize the instrumental stray light, especially that which occurs during off-pointing maneuvers, we made a series of measurements during the BESSY campaigns. By positioning the collimated BESSY beam such that EUV photons still entered SWAP's front aperture, but fell progressively further outside the detector itself, we characterized the behavior of stray light resulting from EUV sources located at different distances outside of SWAP's field-of-view. We collected images of this scattered-light pattern with the beam centered on a range of values from 0 to 32~arcmin outside of the field-of-view in the $y$-direction and in the $x$-direction. (Here $x$ and $y$ refer to the coordinate axis of the detector itself, so $x$ represents translation to the left or right with respect to a plotted image and $y$ translations up or down.) Figure~\ref{fig:straylight_images} presents three representative images obtained during the scattered light tests. Panel~1 shows stray light corresponding to a beam displacement of 8.25~arcmin in the $y$ direction, Panel~2 shows the stray-light pattern when the beam is displaced by 17.25~arcmin in the $y$-direction, while Panel~3 shows that there is relatively little stray light when the beam is displaced by just 8~arcmin in the $x$-direction.

\begin{figure}[ht]
\centering
\includegraphics[scale=0.65, clip = true]{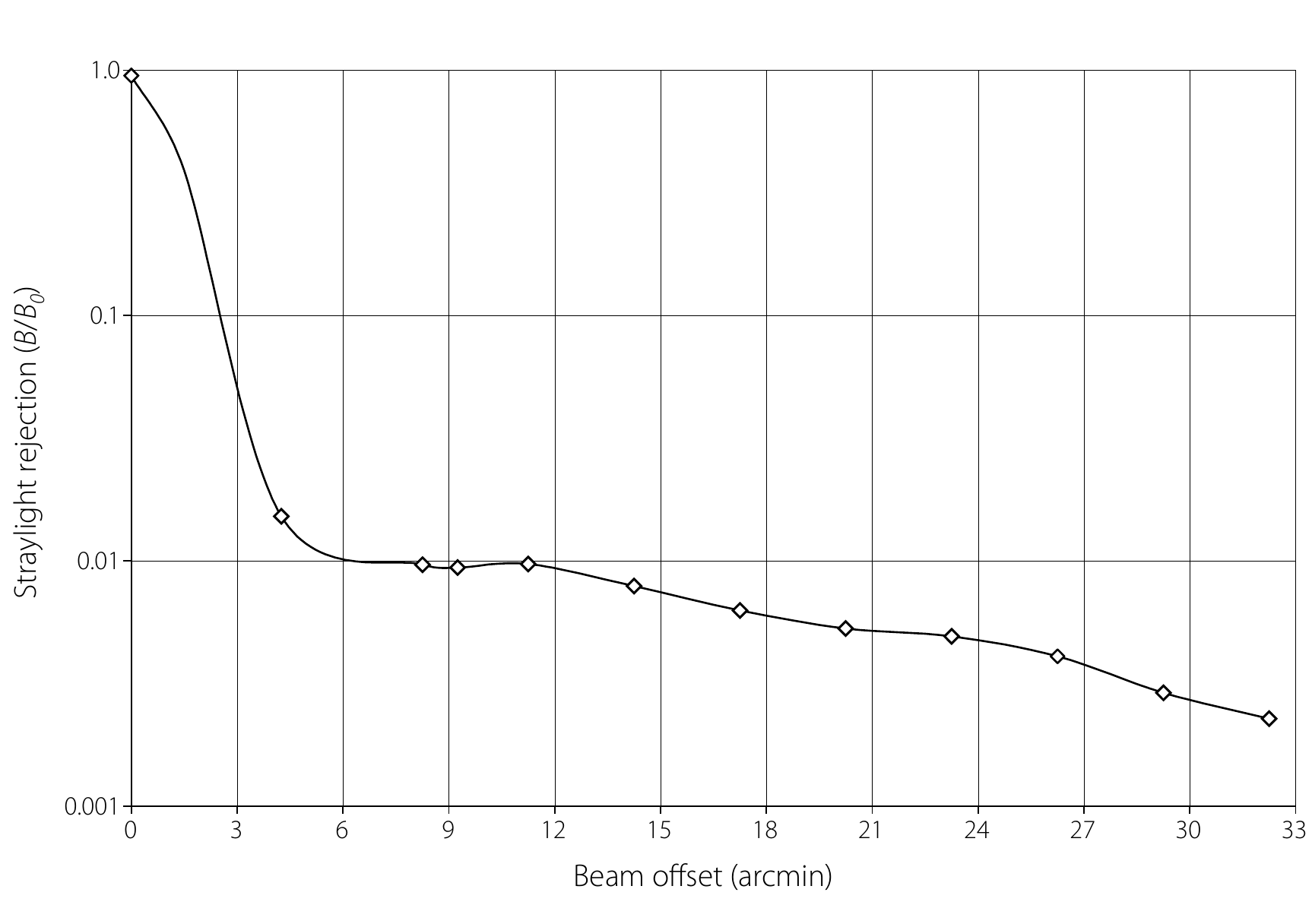} 
\caption{Ratio of integrated intensity for various off-points to intensity of a non-off-pointed beam for various displacements in the $y$-direction.}
\label{fig:functional_straylight}
\end{figure}

Scattered light in the images where the beam was displaced in the $y$-direction clearly showed the characteristic grid-pattern that occurs as a result of the shadowing effect of the FPA filter grid. As a result, we concluded that the source of the stray light must be in front of the rear filter. Figure~\ref{fig:functional_straylight} shows the total observed light flux as a function of beam displacement in the $y$-direction.  The rapid drop in incident flux when the beam is located between zero and three~arcmin occurs as the beam (which is not a single point) leaves the field-of-view, while the relatively flat response beyond that is due to the presence of stray light.  Despite the presence of stray light, incident flux is reduced by approximately a factor of 100 when the light source is outside of the field-of-view.

Ray-tracing simulations conducted using the {\sf ASAP} software package revealed that the likely source of this stray light is grazing incidence reflection caused by the rear filter's acoustic protection baffle.  The baffle edge reflects a part of the beam toward the detector and through the rear filter. The fact that the grid image dimensions appear the same regardless of whether the beam is in or out of the FOV indicates that the grids in both cases are directly imaged without any additional reflections between grid and detector, which further supports the conclusion that the stray light source is the acoustic baffle.

\begin{figure}[ht]
\centering
\includegraphics[scale=0.65, clip = true]{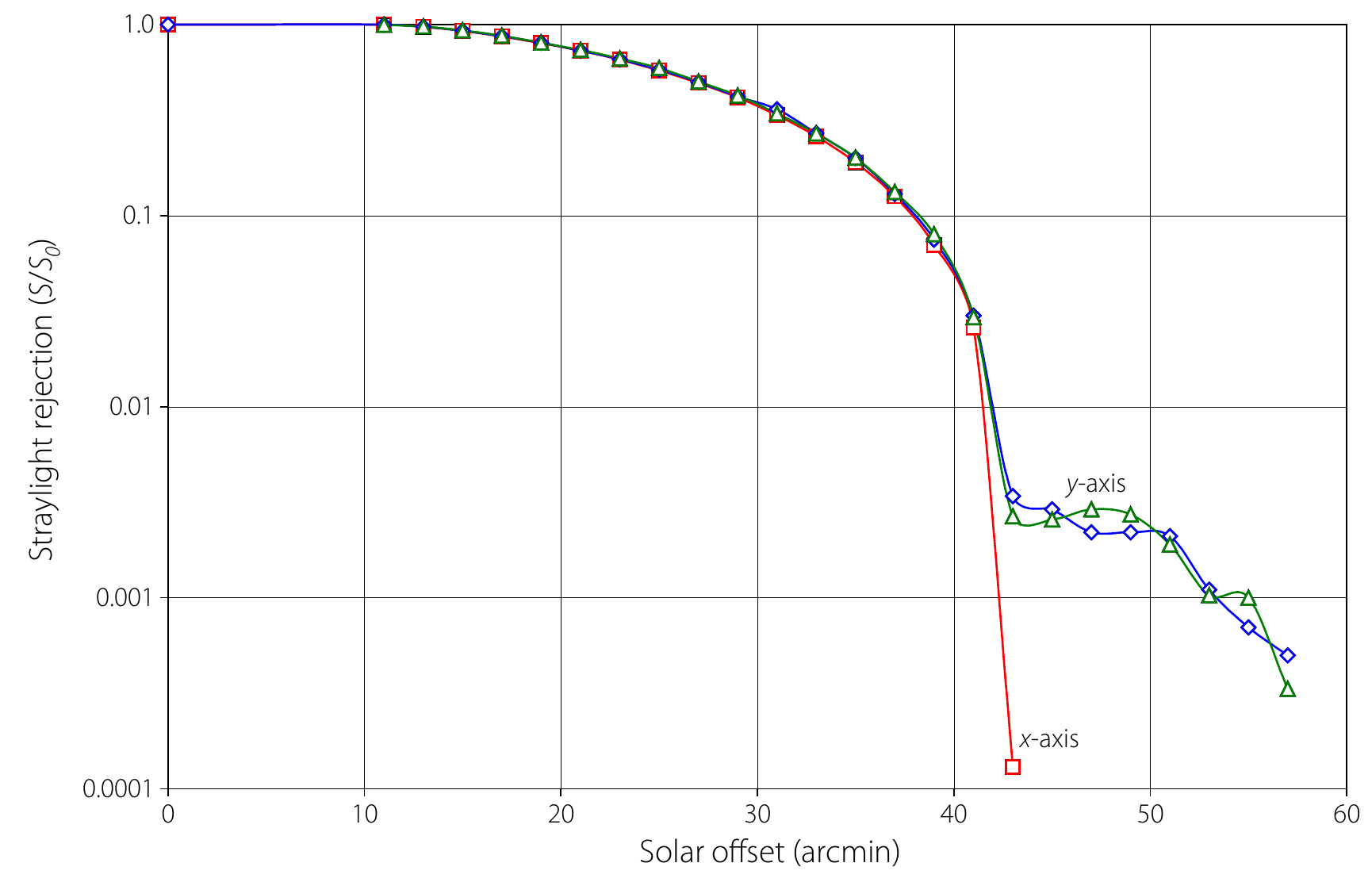} 
\caption{Stray light for solar off-points computed using the {\sf ASAP} ray-tracing software package. The red curve corresponds to off-points in the $x$-direction, while the green and blue curves correspond to the $y$-direction. For comparison, a figure showing the actual in-flight estimation of instrumental stray light appears in the second part of this article \protect\cite{TI_PROBA2_SWAPcalibrationPaper_Halain}.}
\label{fig:raytrace}
\end{figure}

On the other hand, when the beam is displaced in the $x$-direction, scattered light falls off rapidly as soon as the beam leaves the field-of-view (see Figure~\ref{fig:straylight_images}).  Because SWAP is asymmetric in design with respect to $x$ and $y$-axes, such a difference is not surprising.  Because there is nothing in the optical pathway to scatter or reflect light when the beam is displaced in the $x$-direction (as opposed to the acoustic baffle in the other direction) there is almost no stray light originating from sources outside of the field-of-view in the $x$-direction.  Figure~\ref{fig:raytrace} shows ray-trace computations for light sources displaced in both $x$ (red curve) and $y$ (blue and green curves) directions from the field-of-view, confirming that stray light should be expected only when light sources are displaced in $y$. Note that because the ray-trace method is highly sensitive to the number and location of the rays selected for analysis, we performed several simulations of stray light for the $y$-direction using different sets of initial rays. The green and blue curves are the result of two such simulations and they confirm the self-consistency of the result.

Using SWAP's off-pointing capabilities, we have also measured stray light in orbit using a similar procedure. The results of those tests are discussed in detail in the second part of this article \cite{TI_PROBA2_SWAPcalibrationPaper_Halain}.

\subsection{Front Filter Scan}
\label{sec:filter_scan}

In order to characterize the modulation of the signal by the filter grid, we used the BESSY EUV beam to perform a scan of the front filter. To do this, we acquired a series of images between each of which the beam was translated with respect to the instrument so that the incoming beam passed through different locations on the front filter. These translations were made in 30 steps of 0.1~mm in two different, perpendicular, directions with respect to the instrument axes. 

Because only the angle of incidence of the incoming beam\,---\,and not the location at which it enters the telescope\,---\,determines which pixel or group of pixels is illuminated by the beam, holding the beam angle contant means that the beam image appeared on nearly the same pixels in each image of the scan. However, because of ``shadowing'' by the front and FPA filter grids, the image itself changed in appearance as different parts of the beam were attenuated by different parts of the two grids. Depending on the path the beam took through the telescope, the overall image brightness varied by nearly 20\,\%, even though the beam size, intensity, and wavelength during this period were held essentially constant.

In the case of solar images, where incoming photons are not collimated, and light from every part of the Sun falls uniformly on all parts of the entrance filter, the filter grids have only a small effect on the appearance of the image. The front grid causes a net reduction in overall solar signal, but is essentially invisible in the images because photons from a single feature of the Sun can enter the telescope through any location on the front filter, which averages out the effect of any particular gridline. Although the same is true for the rear grid in principle, its location near the detector reduces this averaging effect and, in practice, some grid-related artifacts do appear in solar images from SWAP. However, these artifacts are faint, less than 1\,\% of total signal, and thus are not generally visible in well-scaled images.

\begin{figure}[ht]
\centering
\includegraphics[scale=0.75, clip = true]{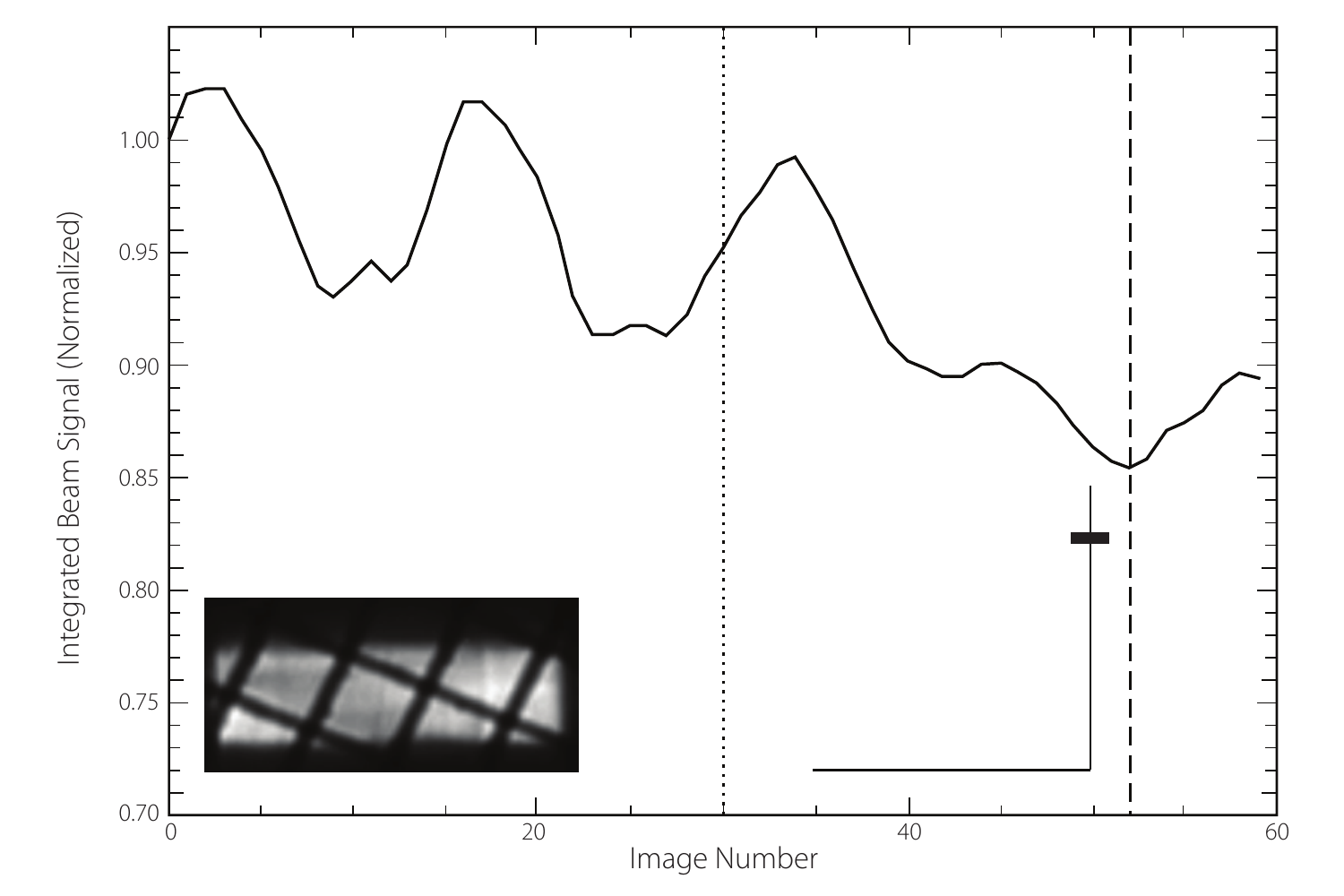} 
\caption{Variation in beam strength as a function of position on the entrance aperture. A sample image showing the filter grid, corresponding to the location at the dashed line, is inset. An inset schematic shows where in the scan the image was obtained. The dotted line indicated the point at which the horizontal scan ended and the vertical scan began. An animated version of this figure showing the evolution of the beam image is available online. The sharply defined grid pattern is generated by the FPA filter, while the much more subtle and diffuse moving shadows are generated as the beam is scanned across the entrance filter grid.}
\label{fig:filter_scan}
\end{figure}

Figure~\ref{fig:filter_scan} shows the variation of the beam strength as a function of position on the entrance aperture during the filter scan. (An animated version of this figure, showing the evolution of the beam image as a result of the filter grids, is available in the online version of this article.) The highly collimated BESSY beam results in grid artifacts that are significantly enhanced from their appearance in solar images. The FPA filter, since it is at the end of the optical path very close to the detector, generates a sharply defined shadow grid pattern, while the entrance filter generates the more rapidly moving, but much less sharply defined grid pattern that also appears in the beam image and movie.


\subsection{Detector Mosaic}

\begin{figure}[ht]
\centering
\includegraphics[scale=0.75, clip = true]{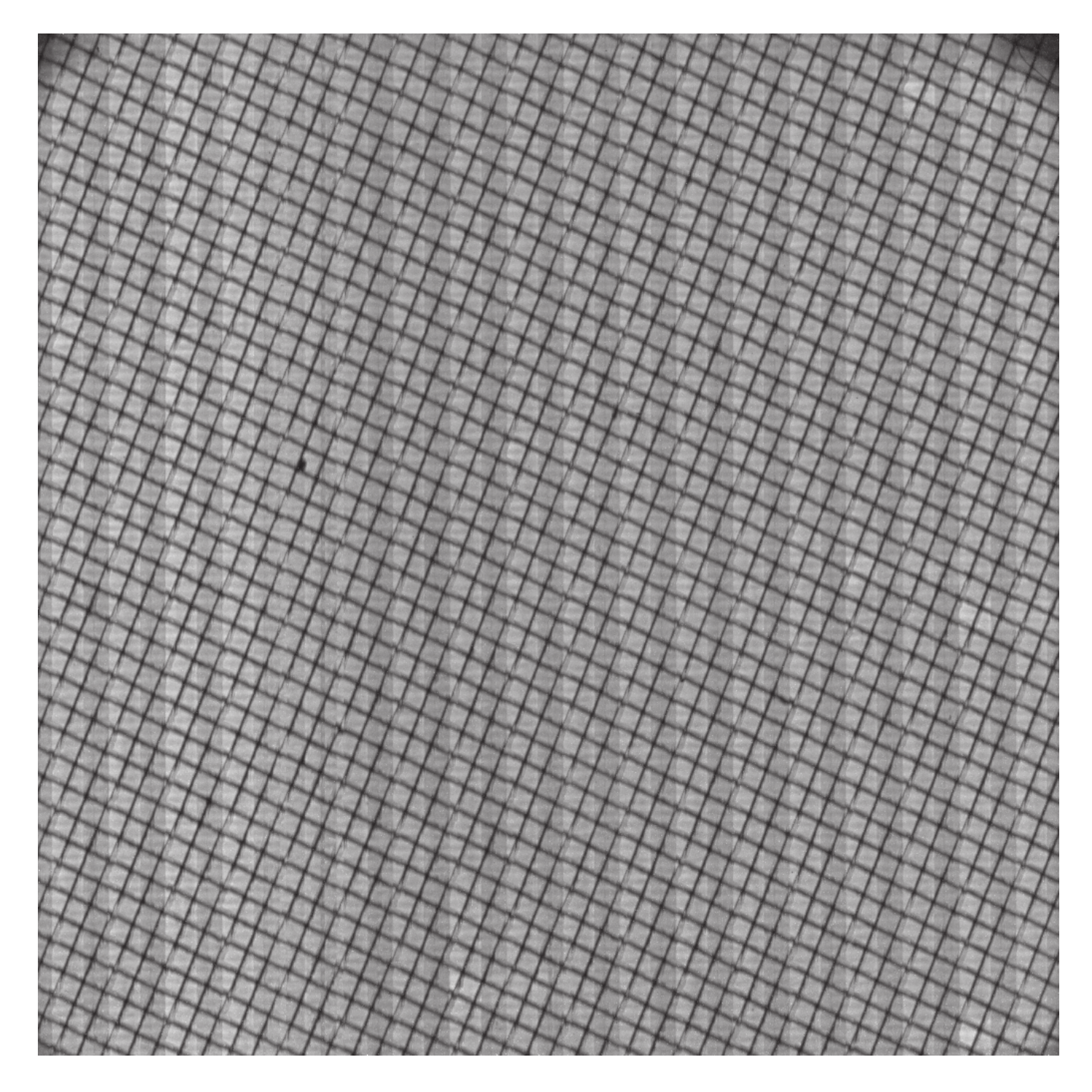} 
\caption{Composite image from filter scans performed at BESSY showing the appearance of the filter grid and vignetting in SWAP images. The grid pattern is the result of shadowing by the nickel support grid on the FPA filter, which has a grid spacing of 0.363~mm. The detector itself is composed of 1024$\times$1024 0.018~mm pixels.}
\label{fig:grid}
\end{figure}

Deriving a flat-field map requires an image in which all detector pixels are (preferentially uniformly) exposed. However, due to the highly collimated EUV beam at BESSY and the small size of the beam spot on the detector, mapping the whole detector was not a trivial task. In lieu of a true flat field, we constructed a detector mosaic using a technique that allowed us to expose each pixel of the detector at least once. By adjusting the incidence angle of the beam slowly over a period of two minutes, we scanned the beam in a vertical band across a strip of the detector as wide as the beam spot. We then adjusted the horizontal angle of the incoming beam, resulting in a translation of the beam horizonally to 19 different positions on the detector, each of which slightly overlapped its neighbors on the left and right. Repeating the  scan procedure in each position resulted in 19 separate vertical scans that we combined to generate a complete map of the detector, which appears in Figure~\ref{fig:grid}.

In this image the highly collimated beam results in the appearance of a sharply defined grid, while in the case of an uncollimated, extended light source like the Sun, the grid is almost invisible. This means that this image is not usable as a pure flat-field correction during image calibration. However, we can nonetheless use the image to estimate the amount of variation in detector response from pixel-to-pixel. Here, we found that, although there is some variation over the detector, it is not a significant source of noise or error. Fewer than 1\,\% of pixels have been identified as malfunctioning completely; in order to improve image compression on board the spacecraft, the values of these poor performers are replaced by the median of their local neighborhood using on board software. These pixels excluded, pixel-to-pixel variation is on the order of a few percent, which is less than the error introduced by other issues such the nonlinearity of pixel response and variability of pixel bias depending on image contents.

\section{Signal Recording}
     \label{SignalRecording} 

The SWAP camera consists of the detector and its surrounding proximity electronics. Together they are soldered to an articulated printed circuit board that is folded into the instrument casing. The main functions of the proximity electronics are the commanding of the detector, the handling of the resulting data and its transmission to SWAP's memory module, the MCPM board in the central spacecraft computer. The proximity electronics also drive two internal LED lamps (used for in-flight calibration measurements) and digitize the signal from temperature sensors inside the telescope body. In the following section, we will restrict the discussion to the details of the SWAP camera and electronics that are relevant to signal recording during normal SWAP science operations.

For its detector, SWAP employs the \emph{High Accuracy Star-Tracker}\footnote{This device is protected by US patent 6225670 and others.} (HAS) active-pixel sensor from FillFactory (Mechelen, Belgium) now part of Cypress. It is a complementary metal-oxide-semiconductor active pixel sensor detector, with 1024$\times$1024~pixels. The HAS datasheet specifies that the three-transistor active pixels, which are 18~$\mathrm{\mu m}$~$\times$~18~$\mathrm{\mu m}$ in size, have a radiation tolerant pixel design and a full-well capacity of 82,000 electrons (with linearity up to 5\,\%). For a pixel illuminated by a point source, 49\,\% of the signal stays within the central pixel, while 9.8\,\% of the signal leaks to each of the direct neighbors $[x \pm 1, y \pm 1]$ and the rest of the signal to further neighbors. The average quantum efficiency (including the fill factor) for the wavelength range of 400~to~720~nm is 40\,\%. We operate the HAS sensor in ``soft reset mode'' (a mode that minimizes read-out noise but can lead to imperfect pixel resets at the end of an integration) by applying 3.3~V as the reset voltage. The antiblooming feature of HAS is left unused by setting the antiblooming voltage to 0~V. A detailed discussion of the electronics and the benefits and drawbacks to the configuration used in SWAP are presented by \inlinecite{Degroof08}.

In SWAP, the sensor surface is covered by a scintillator coating composed of a few layers of small grains that are about five~$\mathrm{\mu m}$ in size (for a total thickness of roughly ten~$\mathrm{\mu m}$). The coating material, commercially known as ``P43'' ($\textrm{Gd}_{2}\textrm{O}_{2}\textrm{S}$ activated by $\textrm{Tb}$), absorbs incoming EUV radiation and re-emits it as visible light (at 545~nm) to which the CMOS--APS is sensitive. Because scintillation pulses decay relatively slowly, this coating creates afterglows that persist for one to three~ms after the initial bright pulse that occurs when an EUV photon strikes the detector coating.  But since the typical integration time of SWAP is ten~seconds, the fluorescence of this coating can be effectively treated as instantaneous. In fact, since the shortest possible integration time for SWAP is one~second, even during the radiometric calibration of SWAP (see Section~\ref{sec:rad_cal}), during which we obtained data using some very short integration times, contributions from these afterglows represent at most a few tenths of one percent of the total acquired signal. Such contributions are below the limits of linearity and noise in SWAP's HAS detector and are thus negligible. Additionally, since the incoming flux is much lower than 300 photons per pixel per second the possibility of phosphor saturation can also be neglected.

One of the main goals of the SWAP project is to demonstrate and test the abilities of CMOS--APS detectors for solar physics in orbit. SWAP's HAS detector offers several advantages over traditional CCD detectors for space-based solar imaging: its radiation resistance, improved noise reduction, ``rolling shutter'' (no mechanical shutter needed), and non-destructive readout mode (the ability to read pixels without resetting them).  As a result of this new readout mode, SWAP has two image acquisition modes: ``Correlated Double Sampling'' (CDS) and ``Double Sampling'' (DS) modes, corresponding to two strategies for reading the signal collected by pixels.

In CDS mode, the sampling mode for almost all nominal SWAP observations, the pixel is read out twice, non-destructively: once at the beginning of the integration process and once at the end of the integration process. The difference between these two values is then computed off-chip after digitizing each sample separately. In this mode the column amplifiers are used in bypass, and the raw signal level (which can be either a bias level following reset or a post-illumination signal level) is sent to the output amplifier, and then to the output for storage and correlated subtraction in the proximity electronics. In this case, since both the bias level and the signal level are from the same reset cycle, their difference not only cancels the fixed pattern noise, but also the temporal readout (kTC) noise.

In DS mode the output signal is computed  from the (analog) difference between the pixel voltage at the end of image integration and the reset voltage immediately following the integration. These two values are then subtracted in the column amplifier, which effectively cancels out most fixed-pattern noise. (The column amplifiers are actually contained in two independent signal buses: one that handles pixels from odd columns and one that handles pixels from even columns. Any offset between these two signal paths can be removed by adjusting a ``column offset'' register.)

As we discussed briefly in Section~\ref{sec:rad_cal}, the response of this signal-processing pipeline deviates from linearity for both very small and very large signals. This is partly because the voltage level on the photodiode is not read out directly (which would destroy whatever image the HAS contains at the time of reading), but rather is read from a source follower in the pixel. When the output is very low, the current source is not constant, and the buffer does not perform linearly. (In DS mode this leads to the so-called ``memory effect'' or ``image lag'', described in detail by \opencite{Degroof08}.) However, the configuration of the HAS detector used in nominal SWAP images, and the fact that most images are obtained in CDS mode, means this is not a concern in most SWAP observations.

For very large signals, there are different steps in the signal processing pipeline that can cause saturation, clipping, and deviations from linearity. When a pixel is integrating, the voltage on the pixel's photodiode becomes more negative as electrons are accumulated. Since the capacitance of the photodiode is voltage dependent, the capacitance increases non-linearly as charge accumulates. According to the instrument datasheet this effect should be smaller than 5\,\% over the measurable range for each pixel, and we have confirmed this with calibration observations both on board and on the ground (see Section~\ref{sec:rad_cal}).

For higher gains, clipping can occur in the on-chip analog amplifier: the measured signal is amplified by a factor 2, 4, or 8, but the total voltage range stays limited to about 2~V, which can cause bright regions to become saturated more quickly than when the gain is set to 1. Amplification is rarely used in normal SWAP observation modes, however, so this type of clipping is of little concern to the average user.

More importantly, clipping also can occur in the analog-to-digital converter (ADC) where any signal coming out of the HAS above 2.04244~V is projected onto the maximum 4095~DN value and any signal below 0.84201~V is projected onto the minimum 0~DN value. In the typical ten-second integration time used in nominal SWAP observations, bright features in the corona do occasionally become saturated, and thus, clipped. However despite the fact that we operate the HAS detector with its antiblooming protection switched off, in SWAP images signal from saturated pixels rarely overflows into pixels beyond their immediate neighbors as often occurs in CCD images, creating large areas of blooming, so saturation in SWAP images is not as much of a concern as it is for CCD-based instruments.

The question of saturation becomes more complicated for images obtained using correlated double sampling mode\,---\,SWAP's normal operating mode\,---\,where the bias frame and the signal frame are digitized independently. For a very bright illumination, the signal frame will be clipped at 4095~DN but the bias frame will still be subtracted from it. This results in an images where the real saturation limit for most pixels is around 3500~DN. It is worth noting, however, that because the bias frame is not recoverable and pixels do not necessarily have exactly the same bias from image to image, it is never possible to be certain that a pixel with a value near 3500~DN in a CDS image is actually saturated or not. However, to optimize image compression, images are typically recoded on board to a maximum value of 3600~DN. Thus any pixel with a value of 3600~DN in a CDS mode image can always be considered saturated.



\section{On Board Data Processing}
     \label{OnboardDataProcessing} 



\begin{sloppypar}
SWAP does not have a dedicated digital processing unit; in order to save on board resources, SWAP images are instead stored and processed by the main platform computer, the \emph{Advanced Data and Power Management System} (ADPMS). The ADPMS is based on a 100 MHz (100 MIPS) AT697 LEON II microprocessor. PROBA2 is the first satellite mission to fly ESA's new high-performance, radiation-hardened processor, which is expected to be used for other future ESA missions such as \emph{Solar Orbiter}. The ADPMS is built as a modular, expandable system with a compact-PCI backplane. For PROBA2, one module, namely the Mass-memory, Compression and Packetization Module (MCPM), provides the SWAP-specific functionality discussed below. The ADPMS, including the MCPM, is doubly redundant on board. One major advantage of the MCPM system for a small mission like PROBA2 is its very small power consumption during nominal operations, which is estimated to be only 2.3~W.
\end{sloppypar}


The MCPM is, first of all, a memory buffer that stores images coming from the SWAP detector. In total three SDRAM memory modules of 2~Gbit each are implemented in the MCPM, of which two modules are used to store SWAP data (\textit{i.e.} 4~Gbit or 512~Mbyte) and the remaining module is used to store the corresponding checksums. At the software level, the 512~Mbytes are divided into two configurable partitions: a small \emph{raw image buffer}, able to store a handful of the latest 1024$\times$1024$\times12$~bit images immediately following acquisition and a larger \emph{processed image buffer} where the images are stored after processing and compression. In its standard configuration, this buffer can store 285 images at a time, although the number of images that can be stored in the buffer is a configurable parameter that is controlled by telecommands sent from the ground. On board, a background process constantly compares the content of the image buffers with associated checksums to detect and\,---\,whenever possible\,---\,correct radiation-induced memory errors.

When SWAP captures an image, it writes it in the raw image buffer of the MCPM. On the ADPMS a software agent called the ``SWAP data manager'' constantly processes SWAP images: it first reads raw picture data from the raw buffer and then processes each image according to the configuration (filtering for bad pixels, compressing the image, textit{etc.}) that was commanded. Finally, the software agent moves the image into a free slot in the processed image buffer.

Before being compressed, discontinuities (typically caused by cosmic rays) are removed and known bad-performing pixels are replaced by on board routines. These two steps help to improve data compression by minimizing the occurrence of JPEG compression artifacts. When the JPEG compression algorithm is selected, the 12-bit pixel values are re-coded to eight-bits before compression. A sophisticated and configurable re-coding algorithm is used to minimize data alteration. Typically, SWAP runs at a high enough cadence that it acquires a few more images on board than can be sent to ground stations. An automatic data prioritization process\,---\,which is supplemented by an observer-configurable priority specification for each image\,---\,identifies potentially interesting solar events to transfer to the ground with highest priority in order to make the best use of limited telemetry bandwidth.


One special feature of PROBA2 has been diverted to improve SWAP's coronal mass ejection (CME) observation program. PROBA2 is a three-axis stabilized platform with agile capabilities. This means that PROBA2 can quickly adjust the pointing of SWAP in any direction. The nearly real-time on board processing of SWAP images allows on board detection of erupting material from the Sun's surface in a set of predefined sectors. If any such event is detected, SWAP's MCPM triggers an off-pointing procedure in order to follow the CME as it propagates away from the Sun. This event detection can also trigger an increase in image cadence or other changes in observational properties.

In principle this exploratory mode increases the total field-of-view of SWAP by a factor of three, but in practice there are a few limitations because of which this mode is rarely used. First, the on board software has difficulty distinguishing CME's from other fast-changing features and artifacts, such as the large number of bright spots that appear in SWAP images due to radiation when the spacecraft passes through the South Atlantic Anomaly (SAA) or pointing error as the result of platform instability. Additionally there have been relatively few eruptions that were intrinsically bright enough to be seen far enough outside of SWAP's nominal field-of-view for the additional operation overhead required for regular use of CME tracking mode to be worthwhile.


\section{Operations}
     \label{Operations} 


SWAP commanding and monitoring is carried out at the PROBA2 Science Center (P2SC) using dedicated software described in detail by \inlinecite{TI_PROBA2_P2SC_Zender}. Although during nominal operations the SWAP returns Sun-centered images obtained at a regular cadence, on board spacecraft operations and limitations as well as commanded observing campaigns can produce images that deviate substantially from the standard data products. In this section we discuss some of the factors that can influence the quality of SWAP data. We also discuss some of the special capabilities and limitations of the instrument itself that have specific bearing on operations.

\subsection{On Board Spacecraft Operations}

Because PROBA2's small size restricted the options for on board control systems, all of the star-trackers necessary for maintaining attitude control are mounted on the top panel of the spacecraft platform. As a result, in order to keep the star field observed by these trackers from being obstructed by the Earth, PROBA2 performs four Large Angle Rotations (LARs) over the course of every orbit. These maneuvers rotate the spacecraft 90~degrees at regular intervals during PROBA2's 98~minute orbit.

As a result, approximately every 25~minutes, the orientation of images from SWAP changes correspondingly. It takes PROBA2 several minutes to carry out the LAR maneuver and restabilize itself afterwards, and any images obtained during one of these maneuvers will be blurred by the motion of the spacecraft. For simplicity, observing operations on board PROBA2 are not suspended during a LAR; instead, LAR-blurred images are detected on the ground and removed from the SWAP data archive during calibration (refer to Section~\ref{DataProducts} for additional detail about ground-based calibration procedures).

Additionally, PROBA2 maintains a fixed alignment with respect to the Ecliptic during almost all nominal operations. As a result, there is often an offset of several degrees between the orientation of a SWAP image and the solar rotation axis (projected onto the image plane). This offset, as well as the regular 90-degree rotations that result from LARs, are corrected during ground-based calibration so that the solar North Pole appears at the top of each image.

There is also another source of image blurring: the reaction wheels that control the orientation of the PROBA2 spacecraft occasionally experience a build-up of angular momentum that must be unloaded in order to maintain spacecraft stability. These episodes normally cause brief periods of platform instability that can also lead to slightly blurred images. Momentum build-up on board PROBA2 is seasonally dependent, so while these episodes can affect as many as 5\,\% of all images during peak periods, for much of the year the effect is limited to only a few images per day.

Another factor that can influence the appearance of SWAP images is the spacecraft pointing itself. Although PROBA2 maintains its Sun-centered pointing during all nominal operations, because the spacecraft and the reaction wheels that control its attitude are relatively small, it is not possible to maintain perfectly precise pointing at all times. As a result, a small amount of drift in the location of the Sun in SWAP images occurs as PROBA2 oscillates slightly around its central pointing axis. This drift varies seasonally and depends a number of factors including changes in the Earth's magnetic field and the star field available to the PROBA2 star trackers. On average this pointing error is less than 30~arcsec, but the variation in pointing over the course of an orbit can be somewhat larger than this average value. During image calibration on the ground, software determines the location of the Sun in SWAP images and corrects for this effect, translating the image so that the Sun appears in the center of the image frame.

Finally, as we discuss in Section~\ref{OnboardDataProcessing}, SWAP's on board software contains routines that can be used to automatically adjust spacecraft pointing to track outgoing CMEs. Although in practice these routines can result in sudden changes in spacecraft pointing, they are not generally activated during nominal operations.

\begin{figure}[ht]
\centering
\includegraphics[scale=0.6, clip = true]{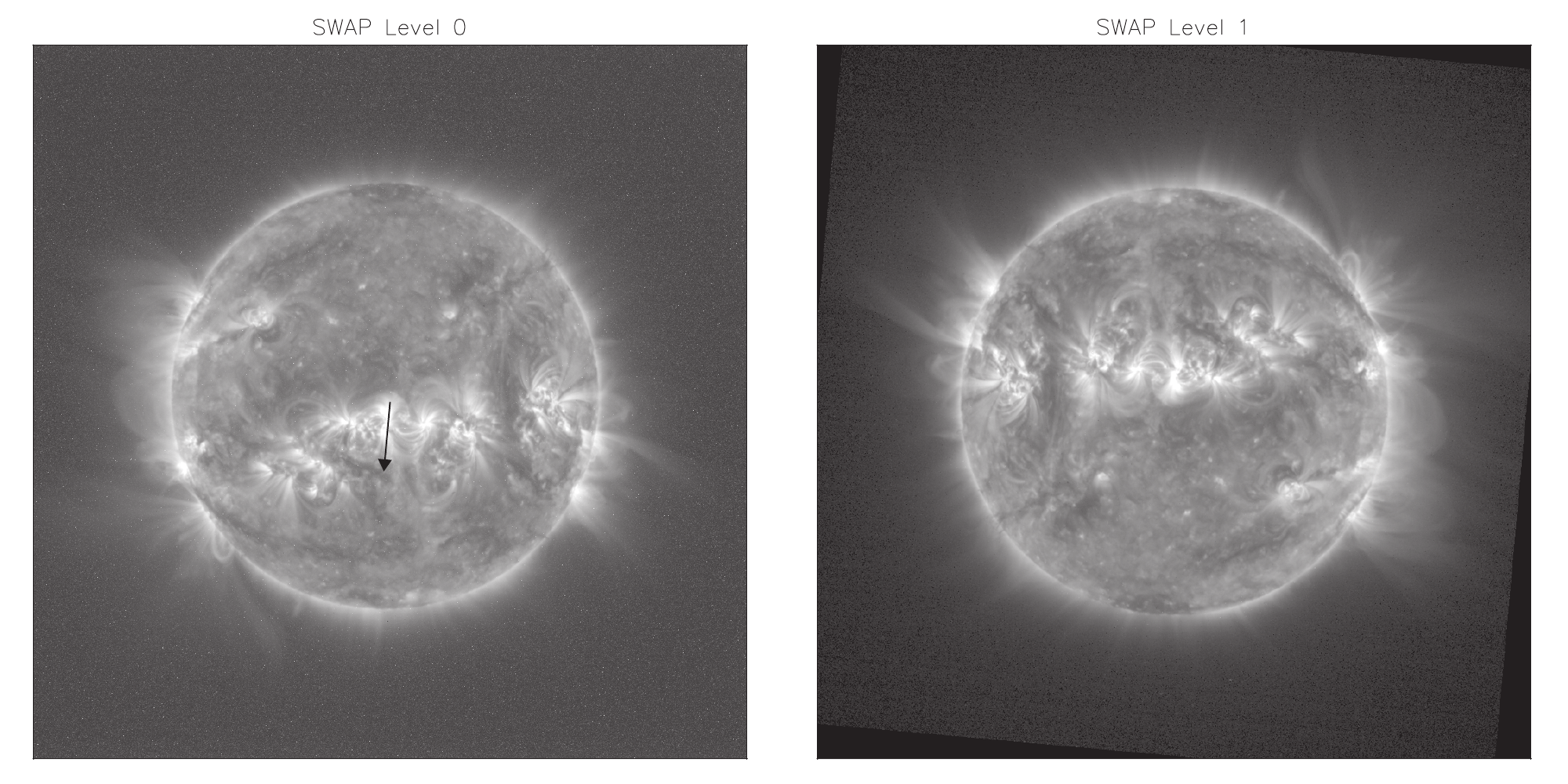} 
\caption{An original SWAP image as observed on board (\textit{e.g.} Level-0, left) and the same image after correction for spacecraft orientation, pointing, and instrumental effects like dark current (\textit{e.g.} Level-1, right). The arrow in the uncorrected image indicated the direction of solar North (North is oriented directly upwards in the Level-1 image).}
\label{fig:lv0_lv1}
\end{figure}

On board housekeeping routines constantly track the spacecraft's location in its orbit, commanded pointing, and orientation. These data are included in as metadata in all SWAP image files, which are distributed in \emph{Flexible Image Transport System} (FITS) format. After image calibration, keywords in the FITS header are updated to reflect the corrected position and orientation of the Sun in SWAP images, but additional keywords maintain information about the original pointing and image orientation for users who wish to track these data. Figure~\ref{fig:lv0_lv1} shows how images are corrected for spacecraft orientation and other instrumental effects. The corrections needed to produce these images are discussed in Section~\ref{DataProducts}.

\subsection{Commanded Spacecraft Operations}

Several types of user-commanded special operations can also affect the appearance of SWAP images as well. Most notably, some observing campaigns make use of SWAP's flexible off-pointing to study large coronal structures that extend beyond the nominal field-of-view, or to extend the field-of-view in the direction of an anticipated eruption. In these cases, the SWAP calibration software automatically detects the commanded off-point and does not translate the position of the Sun to the center of the image frame as it routinely does for nominal images. (Note that user-configurable keywords in SWAP's \textsf{SolarSoft}-based calibration routines allows users to override the default behavior of SWAP's calibration software.)

\begin{figure}[ht]
\centering
\includegraphics[scale=0.6, clip = true]{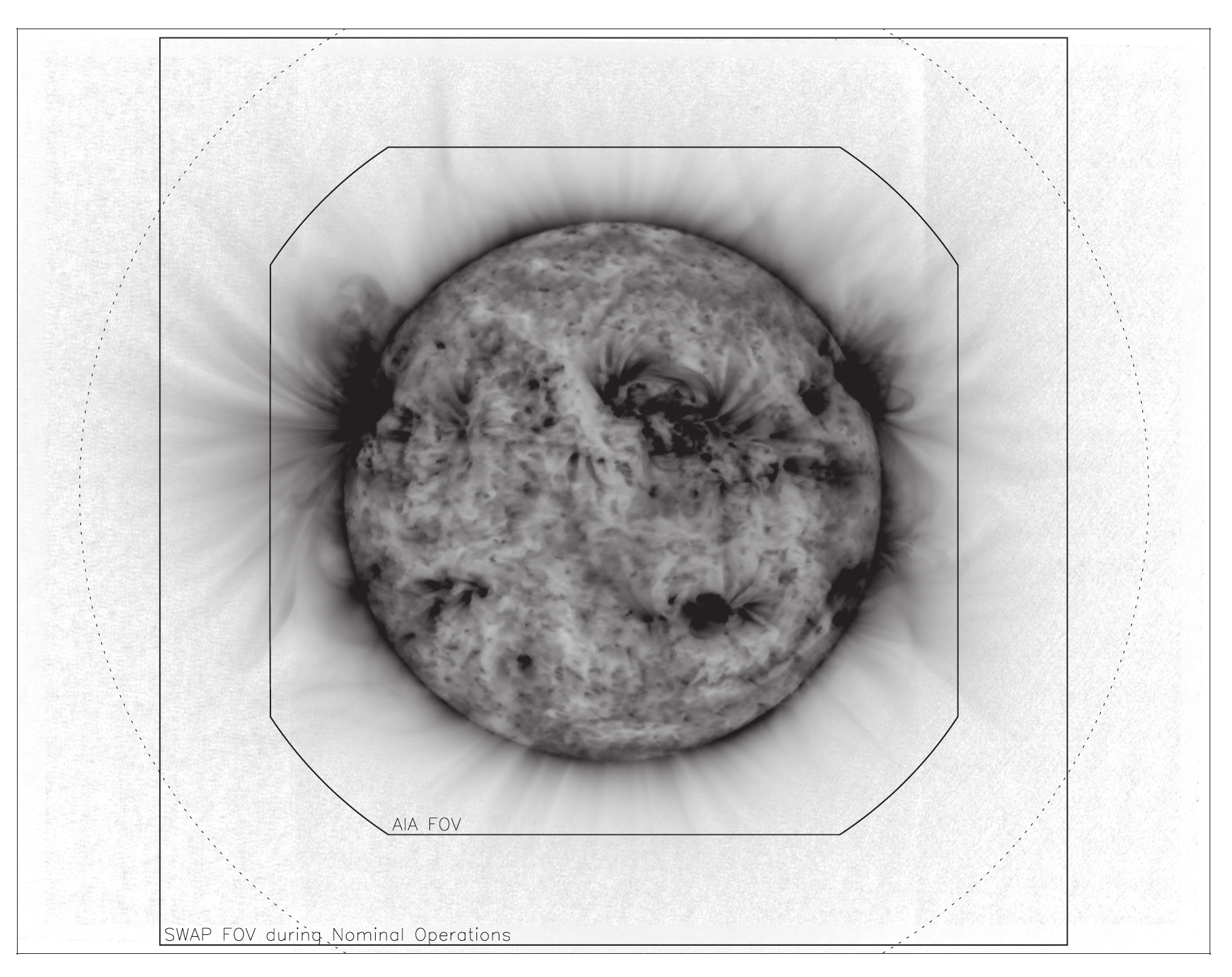} 
\caption{A SWAP mosaic image (seen in inverted color) showing the extent of the EUV corona on 19 September 2011. Overlays on the image show the standard SWAP and AIA fields of view for comparison. The dotted line is at two solar radii, and reveals that many structures extend well beyond the standard SWAP FOV, showing the value of SWAP's mosaic capabilities. (This image has been contrast-enhanced for better display of extended structures.) }
\label{fig:mosaic}
\end{figure}

By off-pointing to several different positions and carefully combining the resulting images, it is possible to generate a mosaic image with an extended field-of-view to show large-scale coronal structures. Such observation campaigns have been useful in filling in gaps between the low corona, which is visible in EUV or white-light eclipse images, and more extended features that are seen in coronagraph images from instruments such as LASCO (\textit{e.g.} \opencite{0004-637X-734-2-114} and \opencite{TI_PROBA2_InnerCorona_Slemzin}). Software for combining several off-pointed images into single mosaic images is provided in SWAP's \textsf{SolarSoft} library. Figure~\ref{fig:mosaic} shows one example of a SWAP mosaic image intended to reveal extended coronal features.

SWAP can also operate in several special modes that affect overall image size, quality, observational cadence, and other properties. During special observing campaigns it is often useful to obtain images with unusual properties in order to improve compression, reduce file size, or track fast-changing features. In order to accomplish this it is possible to rebin images from SWAP to lower resolution (512$\times$512 pixels) or read only a subfield from the whole detector. 

Rebinned images are only one quarter the size of standard images, and as a result, four times as many images can be stored in PROBA2's on board memory buffers and transferred to the ground in the limited spacecraft telemetry bandwidth. For observing campaigns intended to study the rapid evolution of large-scale structures, rebinning allows observers to exchange resolution for higher cadence, a mode that might be especially useful when studying the genesis of large solar eruptions, for example.

Subfields are useful when observers wish to increase cadence, but do not wish to sacrifice image resolution to do so. By reading only a portion of the detector, it is possible to reduce overall image size by ignoring parts of the image that are not of observational interest. However, because of regular changes in spacecraft orientation due to LARs, the sub-region of interest must be redefined every 25~minutes to account for the change in detector orientation, otherwise the region of interest observed will rotate approximately 90~degrees with each LAR.

Finally, it is worth pointing out that calibration campaigns intended to measure degradation in SWAP's response, evolution of dark-current levels, and other image properties occasionally interfere with regular operations. As a result, observers may encounter infrequent data gaps in the SWAP archive. (The results of these calibration campaigns are discussed in the second part of this article, \opencite{TI_PROBA2_SWAPcalibrationPaper_Halain}.)


\section{Data Products and Analysis Software}
     \label{DataProducts} 

Raw SWAP data are converted to images that are distributed in FITS  format in two different data levels: Level-0, uncalibrated images, and Level-1, calibrated and corrected images. In addition, several derivative data products are also available: quick-look images in JPEG format, daily movies, composite Carrington rotation movies, and time-series files containing mean integrated flux which we refer to as SWAVINT (for SWAP Integrated Intensity) files. (All of these data files are available, organized by file level, via the \href{http://proba2.sidc.be/swap/data/}{PROBA2 website}.) SWAP images are also available through the online \href{http://helioviewer.org}{\emph{Helioviewer}} interface.

Level-0 FITS files contain complete image data in raw detector data numbers (DN) with all of the associated metadata needed to generate calibrated images stored as FITS keywords in an image header. Level-1 FITS files are processed by ground-based software to correct for dark current, detector bias (when necessary), flat-field variations, bad pixels, and image artifacts due to radiation exposure in orbit. These images are also corrected so that the Sun is centered in the frame, rotated so that the solar North Pole appears at the top of the image, and rescaled to account for the fact that SWAP's pixels are not perfectly isotropic in size. These images are also time-normalized, so individual pixel values are given in units of $\textrm{DN}\;\textrm{s}^{-1}$. Images that are blurred by LARs or intended for calibration purposes are not converted into Level-1 format; the Level-1 SWAP FITS archive contains only images useful for scientific analysis, while the Level-0 archive contains all images recorded by SWAP.

Although the standard calibration routines produce images that are useful for many scientific purposes, the SWAP calibration routine \textsf{p2sw\_prep.pro} and its associated sub-routines, which are written in the \textsf{Interactive Data Language} (IDL) are distributed publicly via \textsf{SolarSoft} \cite{1998SoPh..182..497F}. These routines are fully customizable via user-defined keywords and can be run on Level-0 images by users who require nonstandard calibration.

\inlinecite{TI_PROBA2_SWAPcalibrationPaper_Halain} provide a detailed discussion of the measurement and evolution of pixel performance, dark current, and other important inputs into the calibration process for SWAP images.

SWAP data files can be obtained via direct download from the PROBA2 website, or interactively using an additional \textsf{SolarSoft} tool that we refer to as the \textsf{SWAP Object}. This tool allows users to download, calibrate, manipulate, and analyze SWAP images inside of an IDL session. Complete documentation and a tutorial for all of SWAP's \textsf{SolarSoft} tools are also included in the SWAP software package itself and additional discussion of how these routines are used in the SWAP data distribution pipeline are presented by \inlinecite{TI_PROBA2_P2SC_Zender}.


\section{Conclusion}
     \label{Conclusions} 

The experience gained from building and operating the SWAP instrument is of particular relevance for developers of the next generation of solar imagers.  SWAP's small size and minimal power requirements have proven that successful science can be carried out even using instruments that operate on resource-poor platforms. SWAP's on board processing capabilities, which both reduce the telemetry required to download data from the spacecraft and optimize the use of limited on board storage by identifying images of maximal interest, provide a foundation for operational planning for the three-bandpass \emph{Extreme-Ultraviolet Imager} (EUI) on the proposed ESA \emph{Solar Orbiter} mission. Given the highly limited telemetry bandwidth available for the mission, on board processing and storage of images is of the utmost importance.  Additionally, EUI, which combines two co-aligned \emph{High Resolution Imagers} (HRI) and one \emph{Full Sun Imager} (FSI) will make use of SWAP's optical design principles as well: one of the HRIs is a two-mirror off-axis telescope with the same bandpass as that of SWAP.

SWAP has also demonstrated that CMOS--APS detectors, which offer advantages of traditional CCD-based detectors, are fully ready for use in scientific remote sensing applications. It is likely that several upcoming solar-imaging missions will make use of similar detectors, and the lessons learned from the SWAP detector can help to inform design and planning for these future instruments.

Finally, with its large field-of-view\,---\,the largest EUV field-of-view currently available from the perspective of the Earth\,---\,and relatively high cadence, SWAP provides images that are simultaneously useful for both science applications and for space-weather monitoring and forecasting. Lessons learned from SWAP have also informed the design of future space-weather monitors, which, given the low overhead required, could easily be integrated into any space-based platform with a small amount of additional space and power.


\appendix


For quick user reference, Table~\ref{instchar_tab1} provides a summary of the main properties of SWAP that are relevant to general users.

\begin{table}
\caption{Instrument Data Sheet}
\label{instchar_tab1}
\begin{tabular}{l c c c}
\hline\hline
Property & Parameter & Value & Units \\
\hline
   Wavelength of peak response 			& $\lambda_{\mathrm{max}}$ & 17.4 & nm \\
   Quantum efficiency of detector 		& $QE$ & 0.45 & n/a \\
   e$^{-}$ per DN conversion factor 	& $C_{e^{-}/DN}$ & 31 & n/a \\
   Geometric aperture area 				& $A_{\mathrm{aper}}$ & 8.55 & cm$^{2}$ \\
   Geometric pixel size 				& $l_{\mathrm{pix}}$ & 0.018 & mm \\
   Telescope focal length 				& $f$ & 1173 & mm \\
   Angular pixel size 					& $l_{\mathrm{pix}}$/$f$ & 3.17 & arcsec \\
   Angular coverage per pixel 			& ($l_{\mathrm{pix}}$/$f$)$^{2}$ & $2.35 \times 10^{-10}$ & str \\
\hline
\end{tabular}
\end{table}


\begin{acks}

SWAP is a project of the Centre Spatial de Li\`ege and the Royal Observatory of Belgium funded by the Belgian Federal Science Policy Office (BELSPO). The work at these institutes and at the K.U.Leuven is supported by PRODEX grant C90193 (SWAP - Preparation to Exploitation), managed by the European Space Agency in collaboration with the Belgian Federal Science Policy Office (BELSPO). We also acknowledge the financial support of the Solar Terrestrial Center of Excellence during the design, testing, and operations of SWAP. We gratefully acknowledge many helpful discussions with  J.-F. Hochedez during the design and testing of SWAP. We acknowledge financial support from the Max-Planck-Institut f\"ur Sonnensystemforschung, Germany, for making the SWAP calibration campaign possible through collaboration with Physikalisch-Technische Bundesanstalt (PTB). We also thank the PTB team of Frank Scholze for their dedication and support during the SWAP calibration campaigns. Part of this work was supported by the German \emph{Deut\-sche For\-schungs\-ge\-mein\-schaft, DFG\/} project number Ts~17/2--1.
\end{acks}


\bibliographystyle{spr-mp-sola}

\end{article} 
\end{document}